\documentclass{aastex}
\usepackage{spr-astr-addons}
\usepackage{graphicx}
\begin{document}
\title{Jet disc coupling in black hole binaries}

\shorttitle{Jet disc coupling}        
\shortauthors{Malzac J.}

\author{Julien Malzac}
\affil{Centre d'Etude Spatiale des Rayonnements, OMP, UPS, CNRS;  9 Avenue du Colonel Roche, 31028 Toulouse, France}

\email{malzac@cesr.fr}




\begin{abstract}

In the last decade multi-wavelength observations have demonstrated the importance of jets in the energy output of  accreting black hole binaries. 
The observed correlations between the presence of a jet and the state of the accretion flow
  provide important  information on the coupling between accretion and ejection processes.
 After a brief review of the properties of black hole binaries,  I illustrate the connection between accretion and ejection through two particularly interesting examples. First, an INTEGRAL observation of Cygnus~X-1 during a 'mini-' state transition reveals disc jet coupling on time scales of orders of hours.
Second, the black hole  XTEJ1118+480  shows complex correlations between the X-ray and optical emission. Those correlations are interpreted in terms of coupling between disc and jet on time scales of seconds or less. Those observations are discussed in the framework of current models.

\end{abstract}



\section{X-ray spectral states and the structure of the accretion flow}\label{sec:xray+geo}

\begin{figure}[t]
\includegraphics[width=\columnwidth]{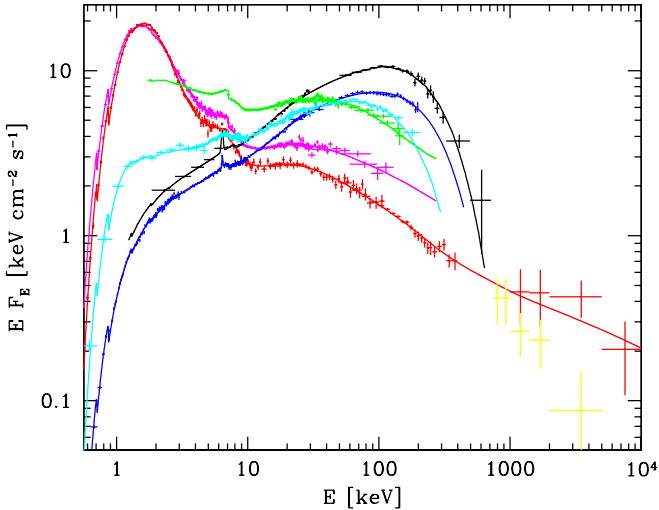}
\caption{Observed spectra of Cygnus X-1 in the LHS (black and  blue), HSS (red, magenta), IMS (green, cyan). The solid curves give the best-fit comptonisation models (thermal in the hard state, and hybrid, thermal-nonthermal, in the other states).  From Zdziarski et al. (2002).
\label{fig:cygx1spectra} }
\end{figure}

Most of the luminosity of accreting black holes is emitted in the X-ray band. This X-ray emission is strongly variable.  A same source  can be  observed with very  different X-ray spectra (see e.g. Zdziarski \& Gierli\'nski 2004).  Fig.~\ref{fig:cygx1spectra} shows various spectra from the prototypical source  Cygnus~X-1 observed at different times. In most sources, there are two main spectal states that are fairly steady and most frequently observed. The occurrence of those spectral states depends, at least  in part, on the variable bolometric luminosity of the source.  At luminosities exceeding a few percent of Eddington ($L_{Edd}$), the spectrum is generally dominated by a thermal component peaking at a few keV which is believed to be the signature of a geometrically thin optically thick disc (Shakura \& Sunyaev 1974).
At higher energies the spectrum is non-thermal and usually presents a weak steep power-law component (photon index $\Gamma \sim 2.3-3$) extending at least to MeV energies, without any hint for a high energy cut-off. This soft power law is generally interpreted as inverse Compton up-scaterring of soft photons (UV, soft X)   by a non-thermal distribution of electrons in a hot relativistic plasma (the so-called "corona"). Since in this state the source is bright in soft X-rays and soft in hard X-rays it is called the High Soft State (hereafter HSS).

\begin{figure}[t]
\centering
\includegraphics[width=\linewidth]{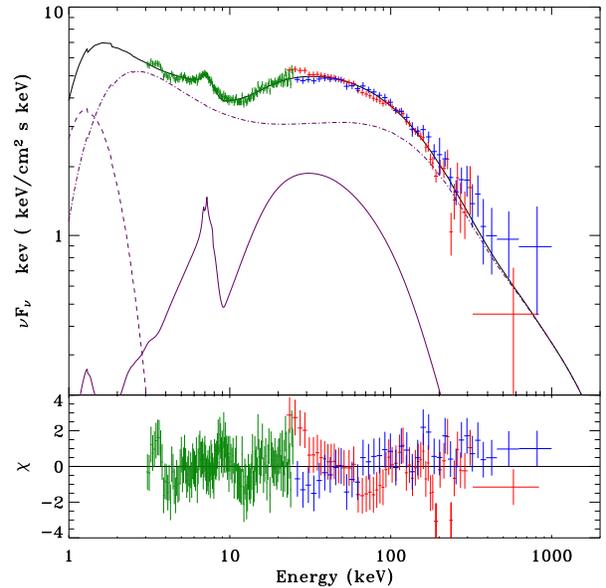}
\caption{Time averaged INTEGRAL spectrum of Cygnus X-1 
during a mini-state transition (intermediate state).
 The data are fitted with the 
thermal/non-thermal hybrid Comptonisation model {\sc eqpair}
 with \emph{mono-energetic} injection of relativistic electrons. 
 The lighter curves show the reflection component (solid), 
 the disc thermal emission (dashed) and the Comptonised emission (dot-dashed).
 The green, red and blue crosses show the {\it JEM-X},
  {IBIS/ISGRI} and {SPI} data respectively. See M06 for details.
\label{fig:spec7980mono} }
\end{figure}

\begin{figure}[tb]
\resizebox{\columnwidth}{!}{\rotatebox{90}{\includegraphics{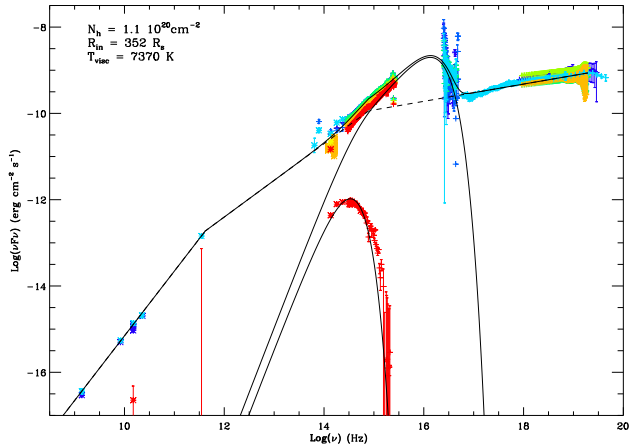}}}
\caption{\label{sed_tout+mod}  Spectral Energy Distribution of XTE J1118+480 during its 2000 outburst (from C03).
\label{fig:sedxte}}
\end{figure}

At lower luminosities (L$<$ 0.01 $L_{Edd}$) , the appearance of the accretion flow is very different: the spectrum can be modelled as a hard power-law $\Gamma \sim 1.5-1.9$
  with a cut-off at $\sim100$ keV. The $\nu F_{\nu}$ spectrum then peaks around a hundred keV. 
 Since  the soft X-ray luminosity is faint and the spectrum is hard, this state is called the Low Hard State (hereafter LHS). LHS spectra are generally very well fitted by Comptonisation  (i.e. multiple Compton up-scatering) of soft  photons by a Maxwellian distribution of  electrons in a hot ($kT_{e}\sim$ 100 keV)  plasma of Thomson optical depth $\tau_{T}$ of order unity. As will be discussed in Sec~\ref{sec:corjetcoup}, the LHS is asssociated with the presence of a compact radio jet that seem to be absent in the HSS. In addition to the dominant comptonisation spectrum there are other less 
prominent spectral features: a weak soft component associated with the thermal emission of a geometrically thin optically thick disc is occasionally detected below 1 keV as observed for instance  in Cygnus X-1 (Balucinska-Church et al. 1995, see Fig.~\ref{fig:spec7980mono}) or in XTE J1118+480 (McClintock et al. 2001, Chaty et al. 2003 hereafter C03, see Fig.~\ref{fig:sedxte}). 
Finally, in both states, reflection features are generally detected in the form of a Fe K$\alpha$ line around  6.4 keV and a Compton reflection bump peaking at about 30 keV, see  Fig.~\ref{fig:spec7980mono}. These components are believed to be produced when the hard X-ray emission of the corona interacts and is reflected by the cold thick accretion disc. They often appear to be broadened through special and general relativistic effects, in which case their origin must be very close to the black hole. These reflection features appear to be weaker in the LHS than in the HSS.

To sumarize, at high luminosities the accretion flow is in the HSS, characterised by a strong thermal disc and reflection component and a weak non-thermal (or hybrid thermal/non-thermal) comptonising corona. At lower luminosity the disc blackbody and reflection features are weaker,  while the corona is dominant and emits through thermal Comptonisation.
Beside the LHS and HSS, there are several other spectral states that often appear, but not always, when the source is about to switch from one of the two main states to the other. Those states are more complex and difficult to define. We refer the reader to McClintock \& Remillard (2006) and Belloni et al. (2005) for two different  spectral classifications based on X-ray temporal as well as spectral criteria and radio emission (Fender 2006).  In general their spectral properties  are intermediate between those of the  LHS and HSS.

\begin{figure}[t]
\includegraphics[width=\columnwidth]{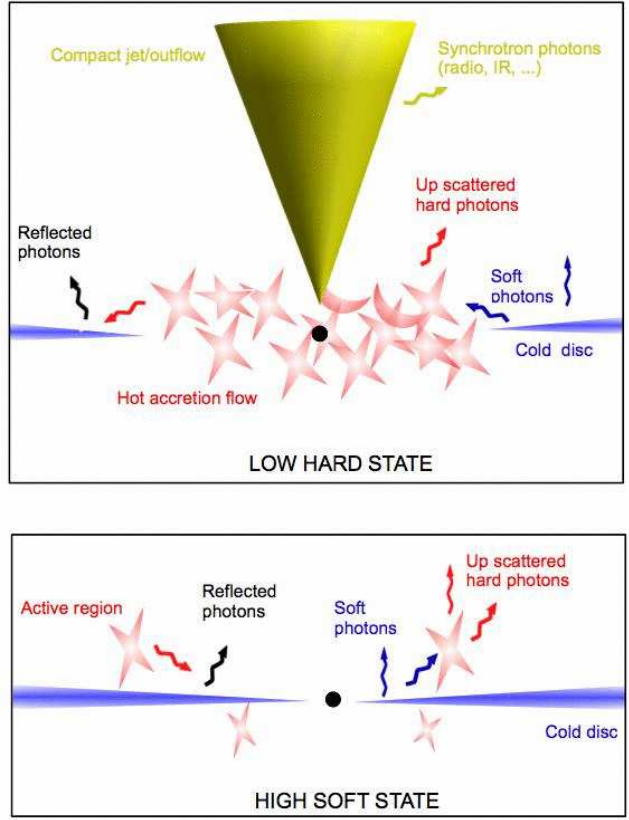}
\caption{Structure of the accretion flow during LHS (top) and HSS (bottom) according to the standard scenario.}
\label{fig:sketch}
\end{figure}

The different spectral states are usually understood in terms of changes in the geometry of the accretion flow.  In the HSS, according to the standard scenario sketched in Fig.~\ref{fig:sketch},  a standard geometrically thin disc extends down to the last stable orbit and is responsible for the dominant thermal emission. This disc is the source of soft seed photons for Comptonisation in small active coronal  regions located above an below the disc. Through magnetic buoyancy the magnetic field lines rise above the accretion disc, transporting a significant fraction of the accretion power into the corona where it is then dissipated through magnetic reconnection (Galeev et al. 1977).  This leads to particle acceleration in the corona.  A population of  high energy electrons is formed which then cools down by up-scattering the soft photons emitted by the disc. This produces the high energy non-thermal emission which in turn illuminates the disc forming strong reflection features (see e.g. Zdziarski \& Gierli{\'n}ski 2004)

In the LHS, the standard geometrically thin disc does not extend to the last stable orbit, instead, the weakness of the thermal features suggest that it is truncated at distances ranging from a few hundreds to a few thousands gravitational radii from the black hole (typically  1000--10000 km). In the inner parts, the accretion flow takes the form of a hot geometrically thick, optically thin disc. A solution of such hot accretion flows was first discovered by Shapiro, Lightman and  Eardley (1976). In these hot accretion flows the gravitational energy is converted in the process of viscous dissipation into the thermal energy of ions. The main coupling between the electrons and the ions is Coulomb collision which is rather weak in the hot thin plasma. Since radiative cooling of the ions is much longer than that of the electrons, the ion temperature is much higher than the electron temperature.  This two temperature plasma solution  is  thermally unstable (Pringle 1976)  but can be stabilised if advection of the hot gas into the black hole (or alternatively into an outflow) dominates the energy transfer for ions (Ichimaru 1977; Rees et al. 1982; Narayan \& Yi 1994, Abramowicz et al. 1996; Blandford \& Begelman 1999; Yuan 2001, 2003). In most of those solutions of advection dominated accretion flows (ADAF), most of the power is not radiated; they are therefore radiatively inefficient.  
The electron have a thermal distribution and cool down by Comptonisation of the soft photons coming from the external geometrically thin disc, as well as IR-optical photons internally generated through self-absorbed synchrotron radiation.  The balance between heating and cooling determines the electron temperature which is found to be of order of 10$^9$ K , as required to fit the spectrum.
The weak reflection features of the LHS are produced through illumination of the cold outer disc by the central source.  
By extension, the hot accretion flow of the LHS  is frequently designated as  "corona" despite the lack of a direct physical analogy with the  rarefied gaseous envelope of the sun and other stars or the accretion disc corona of the HSS. 

\section{An alternative model for the LHS: accretion disc corona}

The presence of the hot accretion disc is not the only possibility to explain the LHS. 
 It was suggested long ago that instead, a real accretion disc corona system similar to that of the HSS but with a pure Maxwellian distribution of comptonising electrons, may reproduce the spectra as well (Bisnovatyi-Kogan \& Blinikov 1976; Liang \& Price 1977). This would imply a cold geometrically thin disc extending down very close to the black hole in the LHS. Unlike in the HSS, this disc does not produce a strong thermal component in the X-ray spectrum because it is too cold. It has a much lower temperature because most of the accretion power is not dissipated in the disc, instead, it is  transported away to power a strong accretion disc corona and the compact jet. 
This, at least in principle, could be achieved either through transport via buoyancy of the magnetic field (Miller \& Stone 2000; Merloni \& Fabian 2002) or through the torque exerted on the accretion disc by a strong magnetic field threading the disc and driving the jet (Ferreira  1997). 

Due to the complexity of the accretion disc corona physics it is not possible to obtain simple analytical solutions predicting the main properties of the corona unless  some parametrization of the energy transfer between the disc and the corona is used (see e.g.  Svensson \& Zdziarski 1994).
However, the appearance of the accretion disc corona does not depend on the details of the energy transport and dissipation mechanisms. Indeed, Haardt \& Maraschi (1991) realised the existence of a strong radiative feedback between the cold disc and the hot corona. A fixed fraction of the power dissipated in the corona in the form of hard X-ray photons,  impinge on the cold disc where it is absorbed, heating the cold disc, and finally reemitted in the form of thermal soft photons. A fraction of those soft photons re-enters the corona providing the major cooling effect to the corona through inverse Compton. Due to this strong feedback from the disc, the cooling rate of the corona scales like the heating rate. The coronal temperature is determined only by the geometry of the accretion disc corona (that controls the fraction of coronal power that returns to the corona in the form of soft photons). For instance an extended corona sandwiching the cold accretion disc, would intercept more cooling  photons from the accretion disc than a patchy corona made of a few compact  active regions covering a small fraction of the disc, and therefore would have a lower temperature and hence a harder comptonised X-ray spectrum. 
During the nineties several groups (Haardt \& Maraschi 1993; Stern et al. 1995 ; Poutanen et al. 1996)
 performed detailed computations of the resulting equilibrium spectra for various geometries. 
They concluded that the corona has to be patchy in order to produce spectra that are hard enough (Haardt et al. 1994; Stern et al. 1995) even when the effects of ionisation on the disc albedo are accounted for (Malzac et al. 2005).
Yet, those accretion disc corona models had an important problem:  the production of an unobserved  strong thermal component due to reprocessing of the radiation illuminating the disc. In the case of an active region emitting isotropically, about half of the luminosity intercepts the disc and the thermal reprocessing component is comparable in luminosity to the primary emission.  Actually, in observed LHS spectra this thermal disc component is so weak that it is barely detectable. Moreover, the amplitude of the reflection features observed in the LHS are lower than what is expected from an isotropic corona by at least a factor of 3, in some cases they are so weak that they are not even detected. 
This led several authors to consider models where the coronal emission is not isotropic. Beloborodov (1999) suggested that the corona is unlikely to stay at rest with respect to the accretion disc. Due to the anisotropy of the dissipation process or simply due to radiation pressure from the disc, the hot plasma is likely to be moving at mildly relativistic velocities. Then, due to Doppler effects, the X-ray emission is strongly beamed in the direction of the plasma velocity. In the case of a velocity directed away from the accretion disc (outflowing corona), the reprocessing features (both reflection and thermalised radiation)  are strongly suppressed. Moreover, due to the reduced feedback from the disc, the corona is hotter, and harder spectra can be produced. 
Malzac, Beloborodov \& Poutanen (2001) performed detailed non-linear Monte-Carlo simulation of this dynamic accretion disc corona equilibrium, and compared the results with the observations. They found that compact active regions of aspect ratio of order unity,  outflowing with a velocity of 30 percent of the speed of light could reproduce the LHS spectrum of Cygnus X-1. 
Moreover since the velocity of the coronal plasma controls both the strength of the reflection features and the feedback of soft cooling photons from the disc, it predicts a correlation between the slope of the the hard X-ray spectrum and  the amplitude of the reflection component. 
Such a correlation is indeed observed in several sources (Zdziarski et al. 1999, 2003) and is well matched by this model.
Recently the accretion disc corona models for the LHS obtained some observational  support, with the discovery of  relativistically broadened iron line in the LHS of GX339-4 (Miller et al., 2006). Such relativistically broadened lines require that disc illumination is taking place very close to the black hole. This observation suggests that, at least in some cases, a thin disc is present at, or close to, the last stable orbit in the LHS.
Incidentally, in the framework of the outflowing corona model, the velocity of the corona required to fit hard state spectra appears to be comparable to  the estimates of the compact radio jet velocity (Gallo et al. 2003). This suggests a direct connection between the corona and the jet as discussed in the next section.

\section{The jet corona connection}\label{sec:corjetcoup}

Multi-wavelength observations
of accreting  black holes in the LHS have shown the presence of
an ubiquitous flat-spectrum radio emission (see e.g Fender 2006), that may
extend up to infrared and optical wavelengths (see Fig.~\ref{fig:sedxte}). 
The properties of the radio emission indicate it is likely  produced
by synchrotron emission from relativistic electrons in compact,
self-absorbed jets (Blandford \& Konigl, 1979; Hjellming \& Johnston 1988). 
This idea was confirmed by the discovery 
of a continuous and steady milliarcsecond 
compact jet around Cygnus X-1  (Stirling 2001).
Moreover, in LHS sources a tight 
correlation has been found between the hard X-ray and radio luminosities, holding over more than three decades in luminosity 
(Corbel et al. 2003; Gallo, Fender \& Pooley 2003).
In contrast, during HSS episodes the sources appear to be
radio weak (Tananbaum et al. 1972; Fender et al. 1999; Corbel 2000), 
suggesting that the Comptonising medium of the low/hard 
state is closely linked to the continuous ejection of matter in 
the form of a small scale jet. 
When the importance of the connection between radio and X-ray emission was realised,
it was proposed that the hard X-ray emission could be in fact synchrotron emission in the jet, rather  than  comptonisation in a hot accretion flow/corona (Markoff, Falke, Fender 2001; Markoff et al. 2003). This model is able to explain  quantitatively the correlation between the X-ray and radio emission, it is also able to reproduce at least roughly the shape of the LHS X-ray spectrum of several sources.  However, it seems that synchrotron emission alone is not enough to reproduce the details of the X-ray spectra. In the most recent version of this model a thermal Comptonisation component was added which appears to provide a dominant contribution to the hard X-ray spectrum (Markoff et al., 2005).  This component is supposedly formed at the base of the jet which forms a hot plasma very similar to an accretion disc corona. 
In the context of accretion disc coronae/hot disc models the correlation between X-ray and radio emission, simply tells us that the corona and the compact jet of the LHS are intimately connected. 
A strong corona may be necessary to launch  a jet and/or could be  the physical location where the jet is accelerated or launched (Merloni \& Fabian 2002).  
K\"{o}rding et al. (2006) show that the observed X-radio correlation can be reproduced provided that the 
hot flow/corona is radiatively inefficient (i.e. luminosity scales like the square of the mass accretion rate) and that a constant fraction of the accretion power goes into the jet (i.e. jet power scales like mass accretion rate). 
However there is presently no detailed model to explain how the physical connection between jet and corona works.  A possible basic explanation was proposed by Meier (2001) for ADAF like accretion flows and later extended to the case of accretion disc coronae by Merloni and Fabian (2002).  It goes as follows. Models and simulations of jet production (Blandford \& Znajek 1977;
Blandford \& Payne 1982, Ferreira 1997)  indicate that jets are driven 
by the polo\"{i}dal component of the magnetic field. If we assume that the magnetic field is generated by dynamo processes in the disc/corona, the strength of the poloidal component is limited by the scale height of the flow (Livio, Ogilvie \& Pringle 1999; Meier 2001; Merloni \& Fabian 2002). 
Therefore geometrically thick accretion flows should be naturally more efficient at launching jets.

\begin{figure}[t]
\includegraphics[width=\linewidth]{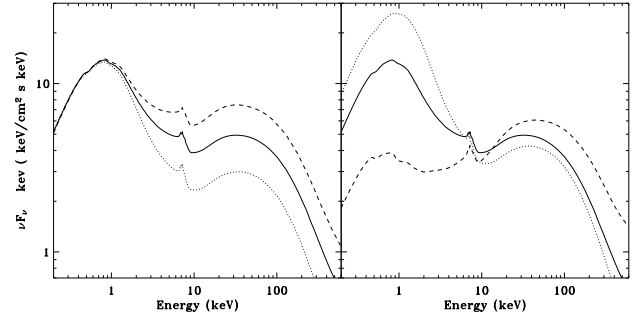}
\caption{Modelling the spectral variability  of Cygnus X-1 during its mini-state transition. Left panel: effect of varying the heating rate in the corona ($l_{h}$ parameter of the {\sc eqpair} model) by a factor of 2. 
Solid curve: unabsorbed best-fit model of the time-averaged spectrum shown in Fig.~\ref{fig:spec7980mono} ($l_{h}=8.5$); Dotted curve  $l_{h}=5.7$; 
 Dashed curve: $l_{h}=11.9$. 
 Right panel: effect of varying the soft photons flux in the hot corona ($l_{s}$ parameter of the {\sc eqpair} model) by a factor of 8 at constant heating rate. 
 Solid curve: unabsorbed best-fit model  
 ($T_{\rm disc}=0.3$ keV; $l_{\rm h}/l_{\rm s}=0.85$). 
 Dotted curve: $T_{\rm disc}=0.357$ keV and $l_{\rm h}/l_{\rm s}=0.42$.
 Dashed curve: $T_{\rm disc}=0.212$ keV and $l_{\rm h}/l_{\rm s}=3.4$. From M06.}
 \label{fig:interp} 
\end{figure}

  \begin{figure}[th]
  \centering
  \includegraphics[width=\linewidth]{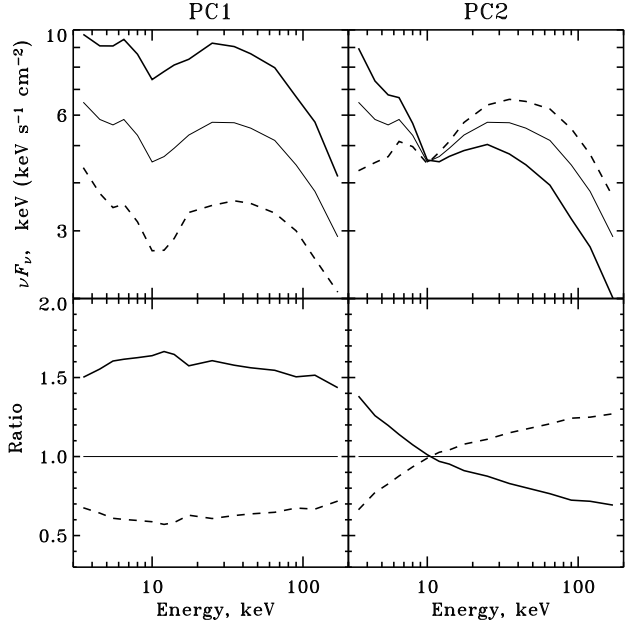}
      \caption{The 2 principal components of variability of Cygnus X-1 during its mini state-transition. 
      The upper panels illustrate the effects of each component on the shape 
      and normalisation 
      of the spectrum: time average spectrum (light line) 
      and spectra obtained for the maximum and minimum observed values 
      of the normalisation parameter.
      The bottom panels show the ratio of spectra obtained for the maximum and 
      minimum normalisation to the average one. See M06 for details.
              }
         \label{fig:pcatot}
   \end{figure}
      \begin{figure}[tb]
   \centering
 \includegraphics[width=\linewidth]{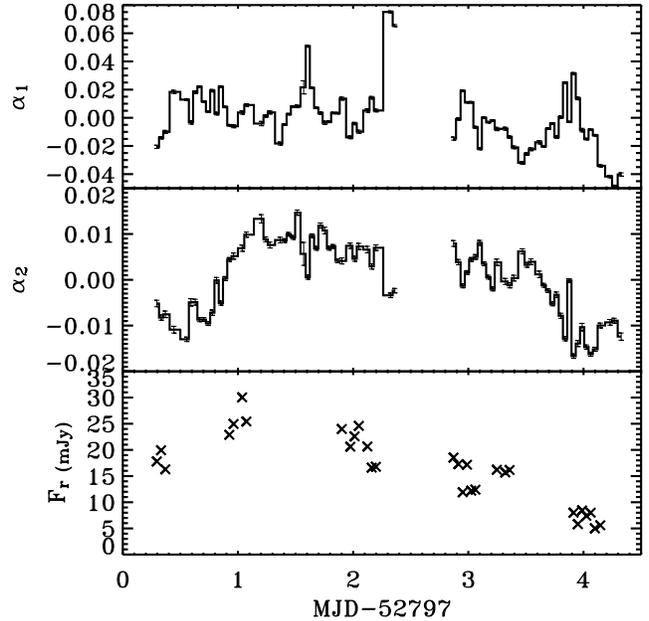}
      \caption{ Spectral variability of Cygnus X-1 during a mini-state transition. Evolution of $\alpha_1$, the amplitude of PC 1,  tracer of the luminosity (top); $\alpha_2$ , amplitude of PC 2, tracer of the hardness (middle); and radio light curve (bottom) during the observation. From M06}
         \label{fig:pcavar}
   \end{figure}

     \begin{figure}[tb]
        \centering
 \resizebox{\columnwidth}{!}{\includegraphics{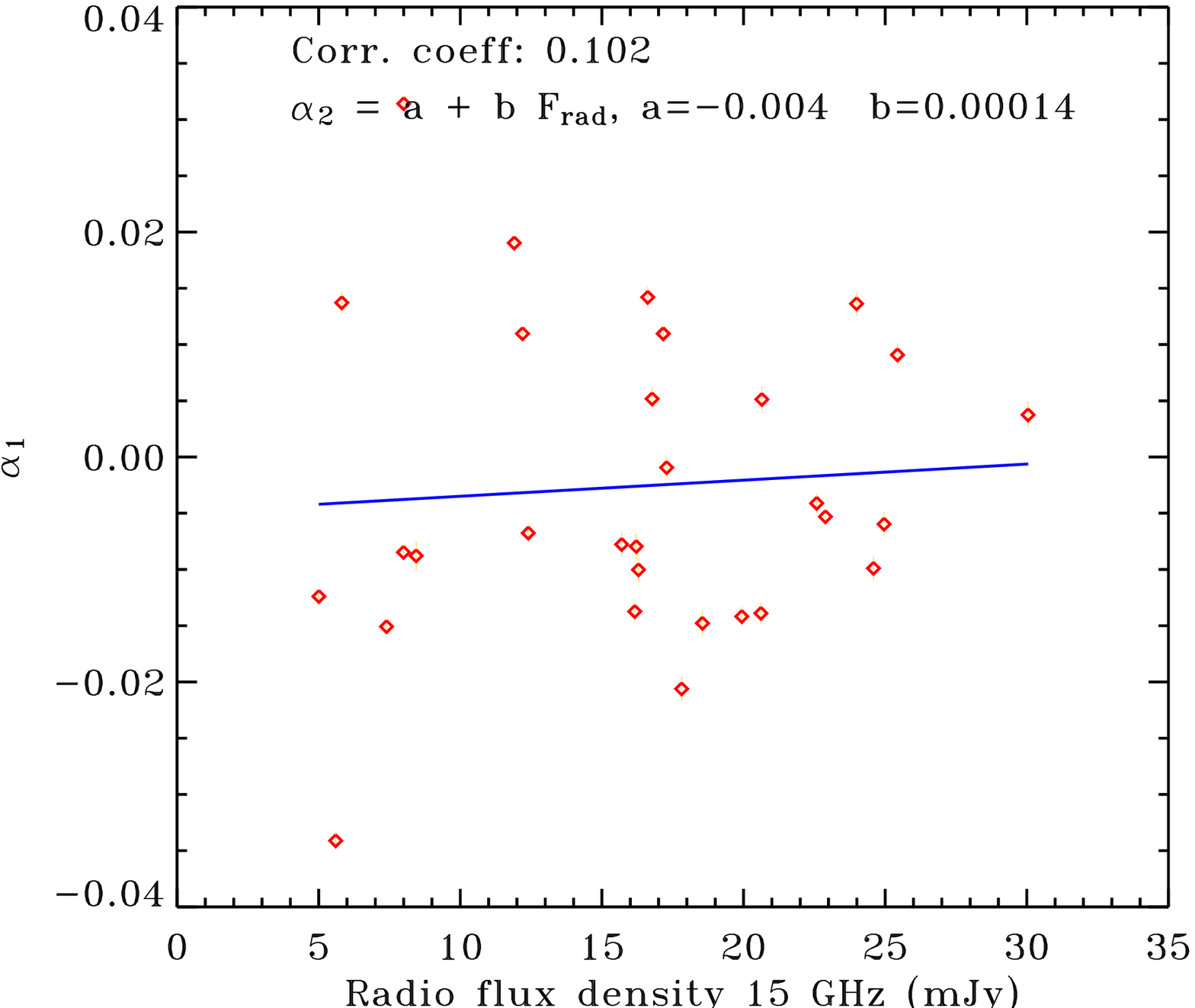}\includegraphics{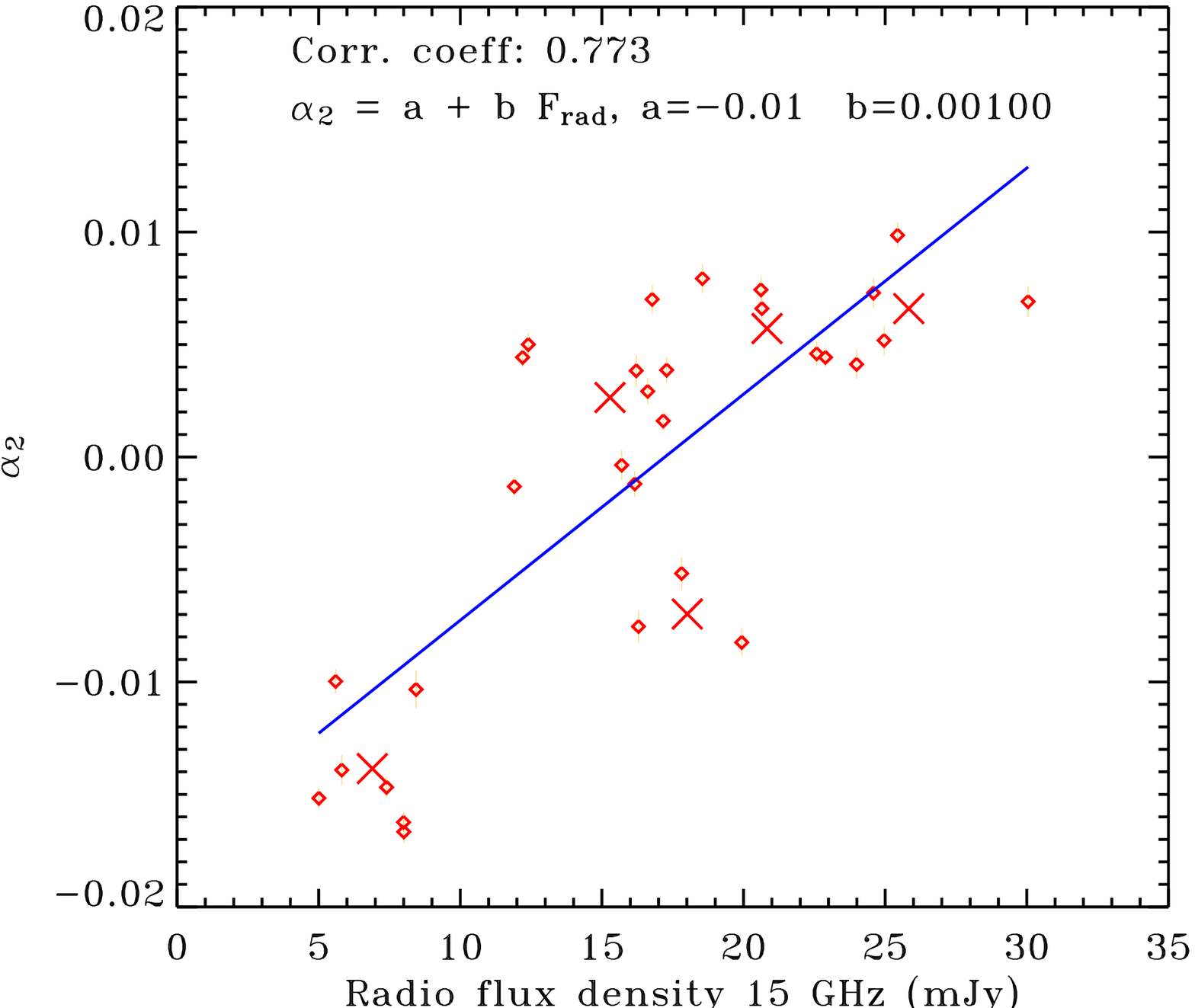}}
      \caption{The amplitude of PC 1, $\alpha_{1}$ (tracer of the luminosity, left panel) and PC 2,  $\alpha_{2}$ (tracer of the hardness, right panel) 
      as a function of the radio flux (diamonds). In both panels, the best linear fits are shown by the solid lines. 
The crosses indicate the time average over each of
the five periods of nearly continuous radio coverage 
 (see Fig ~\ref{fig:pcavar}).  While there is no convincing correlation between the
  radio flux and $\alpha_{1}$, the radio flux is correlated 
  to $\alpha_{2}$ at highly significant level. (From M06)}  
      \label{fig:radal}
   \end{figure}

\section{Jet disc coupling during a mini-state transition in Cygnus X-1}

As discussed above, in the LHS, there is a strong connection between hot corona 
and outflow. On the other hand, during state transitions, the situation seems to be different. 
It appears indeed that  the jet  is then connected to the cold accretion disc rather than the corona. For illustration I summarise below the results of  Malzac et al. (2006, hereafter M06)  reporting a coordinated observation of the source Cygnus X-1  in the hard X-rays  (INTEGRAL)  and at radio wavelengths (Ryle telescope). Cygnus X-1 is the prototype of black hole candidates. 
Since its discovery in 1964 (Bowyer et al. 1965), it has been intensively observed by 
all the high-energy instruments, from soft X-rays to $\gamma$-rays. 
It is a persistent source most often observed LHS switching occasionally  to the HSS.
There are also Intermediate States (hereafter IMS)
in  which the source exhibits a
relatively soft X-ray spectrum ($\Gamma \sim 2.1-2.3$) and a
moderately strong soft thermal component (Belloni et al. 1996; Mendez \& van der Klis 1997, see Fig.~\ref{fig:cygx1spectra}). 
During the 4 days long  observation of M06, the source was in an IMS,  as demonstrated by the appearance of the  INTEGRAL time-averaged spectrum shown in Fig.~\ref{fig:spec7980mono}.  This spectrum is fitted with a thermal/non-thermal hybrid Comptonisation model 
({\sc eqpair} model, see Coppi 1999; Gierli\`nski et al 1999, hereafter G99; Zdziarski et al. 2002, 2004) 
 with mono-energetic injection of non-thermal electrons. 
 The resulting best fit parameters are intermediate between what is usually  found in the LHS and HSS.
  The unabsorbed best fit model spectrum is shown on Fig.~\ref{fig:interp}.
 In order to study the spectral variability of the source during this
observation, M06 produced light curves in 16 energy 
bands ranging from 3 to 200 keV
with a time resolution of the duration of a science
window (i.e. $\sim$ 30 min). 
The light curves  exhibit a complex 
and strong broad band variability of the spectra as well as the overall flux.
During the 4 day long observation 
the broad band (3--200 keV) luminosity varied by up to a factor of 2.6 and 
the source showed an important spectral variability. 
A principal component analysis demonstrates that most of this variability
occurs through 2 independent modes shown in Fig.~\ref{fig:pcatot} and~\ref{fig:pcavar}. 
The first mode (hereafter PC 1) consists in 
 changes in the overall luminosity on time scale of hours
 with almost constant spectra (responsible for 68 \% of the variance). 
 M06 interpret this variability mode as variations of the dissipation rate in the corona, 
 possibly associated with magnetic flares. 
 The second  variability mode (hereafter PC2) consists in a pivoting of the spectrum around
  $\sim$10 keV (27 \% of the variance). The two spectra obtained for the minimum  and maximum values of $\alpha_{2}$
 parameter controlling the amplitude of PC 2 are reminiscent of 
 the canonical LHS and HSS spectra. The evolution of  $\alpha_{2}$ demonstrates the spectral evolution during the observation: 
  initially soft, the spectrum hardens in the first part of the observation and then softens again  (see Fig.~\ref{fig:pcavar}).
This pivoting pattern is strongly correlated with the radio (15 GHz) emission:
 radio fluxes are stronger when the {\it INTEGRAL} spectrum is harder (see Fig.~\ref{fig:pcavar} and~\ref{fig:radal})  On the other hand, there is no hint of a correlation with the flaring mode as can
     be seen in the left panel of Fig.~\ref{fig:radal}. 
     In other terms, the radio emission is strongly correlated with the hardness 
     and apparently unrelated to 2 to 200 keV luminosity.

  The pivoting mode actually represents
 a 'mini' state transition from a nearly HSS to a nearly LHS,
  and back. 
 Compilations of LHS and HSS spectra
   indeed suggest that the spectral transition between LHS and HSS occurs
    through a pivoting around 10 keV (see e.g. Fig. ~\ref{fig:cygx1spectra}).
The transition from LHS to HSS is known to be 
associated with a quenching of the radio emission (Corbel et al. 2000;
Gallo, Fender \& Pooley 2003). 
When, during the observation, the source gets closer to the HSS 
the spectrum softens and simultaneously the radio flux decreases producing the observed correlation between radio flux and hardness. 

 M06 tried to reproduce the 2 variability modes by varying the parameters  of the
 hybrid thermal/non-thermal Comptonisation models shown in Fig.~\ref{fig:spec7980mono}.
 As shown in  the left panel of Fig.~\ref{fig:interp} it is possible to produce
   variations in luminosity by a factor comparable to what is observed
 and little spectral changes in the {\it INTEGRAL} band  by varying the heating rate in the corona (or equivalently the coronal 
 compactness $l_{\rm h}$) by a factor of 2.
 In this context the flaring mode would correspond to variations of the dissipation rate
  in the corona possibly due to magnetic reconnection.
  This variability mode seems to be a characteristic of the HSS (Zdziarski et al. 2002).
  As we show here, it also provides a major contribution
   to the variability of the IMS.  
  Regarding the pivoting mode, it can be
 produced by changes in the flux of soft cooling photons (or soft photon compactness parameter, $l_{s}$ ,  in {\sc eqpair}), at constant
  dissipation in the hot phase. 
M06 performed simulations assuming that the accretion disc
  radiates  like a blackbody i.e.
  its flux $F_{disc} \propto l_{\rm s} \propto T_{\rm max}^4$ and constant $l_{h}$.
   For an increase of the disc temperature by a factor of 1.7, the disc luminosity 
   grows by a factor of 8. As the disc flux also corresponds
    to the soft cooling photon input in the corona and the heating ($\propto l_{\rm h}$)
     is kept constant, this leads to a steepening of the spectrum
      with a pivot around 10 keV of similar amplitude as in PC 2 
  (see Fig.~\ref{fig:interp}). 
   For the 1996 HSS, G99 found a ratio $l_{h}/l_{s}\sim0.3$ while
   in the LHS, $l_{h}/l_{s}$ ranges between 3.5 to 15 (Ibragimov et al. 2005).
   The range of $l_{h}/l_{s}$ (0.4--3.4) required to reproduce the observed amplitude
    of the pivoting mode matches almost exactly the intermediate range between the HSS 
    and the lower limit of the LHS.  The source initially in a (quasi) HSS evolved
   toward the LHS but as soon as it was reached, it went back toward the HSS.
    In this interpretation the radio versus X-ray hardness correlation suggest that the jet power is anti-correlated with the disc luminosity and unrelated to the coronal power.
    The reason for this anti-correlation is most probably  that  during state transition are associated with a redistribution of the available accretion power between the compact jet and the cold accretion disc, in the standard scenario described in Sec.~\ref{sec:xray+geo} this redistribution of accretion power could occurs because the jet  shrinks as the inner radius of the outer discs moves closer to the black hole.    
    
\section{Optical/X-ray correlations as a signature of jet-disc coupling in XTE J1118+480}
\label{}

\begin{figure}[!tb]
  
  \includegraphics[width=\columnwidth]{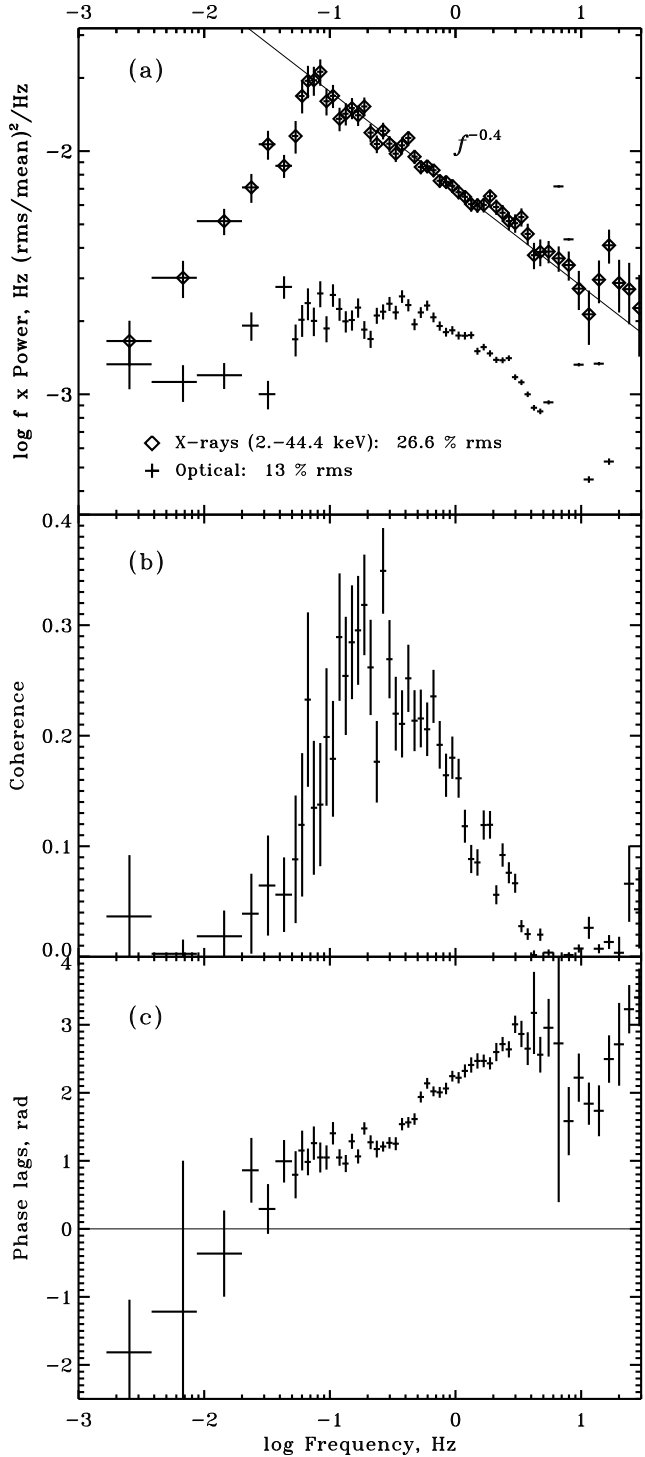}
   \caption{The optical/X-ray correlations  of XTE J1118+480  in the Fourier domain.
 {\bf a)} X-ray and optical power spectra. The counting noise was
subtracted.
 {\bf b)} X-ray/optical
coherence. {\bf c)} phase-lags as function of
Fourier frequency. A positive lag implies 
that the optical is delayed with respect to the X-rays. From M03}
  \label{fig:corxofour}
  \end{figure}

The  X-ray nova XTE J1118+480, was discovered by the Rossi X-Ray Timing Explorer
({\it RXTE}) All-Sky Monitor ({\it ASM}) on 2000 March 29 
(Remillard et al. 2000).
 The optical spectrophotometry proved a low mass X-ray binary system
 containing a black hole of at least 6  solar masses
(McClintock al. 2001a).
 The interstellar extinction towards
  the source is exceptionally low (Garcia et al. 2000).
 This fact allowed an unprecedented wavelength coverage
 (McClintock et al. 2001b;
 Hynes et al. 2003 hereafter H03; C03 and references therein).
 The resulting spectral energy distribution is displayed in Fig.~\ref{fig:sedxte}.
In  the radio to optical bands,  
a strong non-thermal component was associated with synchrotron
emission from a powerful jet or outflow (Fender et
al. 2001).  In the
optical to EUV bands the spectral energy distribution 
 is dominated by a thermal component from the accretion
disc. The X-ray emission consists of a typical powerlaw spectrum with photon
index $\Gamma \sim 1.8$. 
During the whole outburst duration, the X-ray properties of 
the source, as well as the presence of strong radio emission,
were typical of black hole binaries in the hard state.
The total mass accretion rate onto the black hole can be estimated
from the observed luminosity of the cold disc component which in the
case of XTE J1118+480 can be estimated from fits to the optical, UV
and EUV spectra (see e.g. C03).  The result depends on
several assumptions regarding the geometry and the physical mechanisms
for energy dissipation in the disc and the hot comptonising medium.
However, for reasonable parameters, the estimated accretion rate is
much larger than the observed bolometric luminosity (by at least a
factor of ten) and therefore the source is radiatively inefficient. 
 The important issue, however, would be to determine
whether the missing accretion power escapes the system in the (low
radiative efficiency) jet or in other forms of non-radiative losses,
such as a slow wind, or large scale convective motions, or advection
into the black hole.  The answer to this question resides in the exact
determination of the jet kinetic power.  Unfortunately, there are
major uncertainties in this determination, mainly because the jet
radiative efficiency is not known.  The jet is expected to be a poor
radiator because most of the energy is lost in adiabatic
expansion. Thus, although the radiation from the jet represents a
small fraction of the bolometric luminosity the jet could dominate the
energetics.  For the case of XTE~J1118+480, typical efficiency of
$\epsilon_{\rm j}\sim 0.01$ would already imply that the total jet
power dominates over the X-ray luminosity. 
Interestingly, fast optical and UV photometry allowed by the weak extinction, revealed a rapid optical/UV
flickering presenting complex correlations with the X-ray variability (Kanbach et al. 2001, hereafter K01; H03). 
This correlated variability cannot be caused by reprocessing 
of the X-rays in the external parts of the disc.
Indeed, the optical flickering occurs on average on shorter
time-scales than the X-ray one (K01), and reprocessing models fail to 
fit  the complicated shape of the X-ray/optical cross correlation 
function (H03).
Spectrally, the jet emission  seems to
 extend at least up to the optical band (C03), 
although the external parts of the disc may provide an
important  contribution to the observed flux at such
wavelengths.
The jet  activity is thus the most likely explanation for the rapid
observed optical flickering. For this reason, 
the properties of the optical/X-ray correlation 
in XTE J1118+480 might be of primary importance for the understanding 
of the jet-corona coupling and the ejection process.
 The simultaneous optical/X-ray observations are described at length in
a number of papers (K01; Spruit \& Kanbach 2001; H03; Malzac et
al. 2003, hereafter M03). As discussed in these works, the observations are very challenging
 for any accretion model. The most puzzling pieces of evidence 
are the following:   
(a) the optical/X-ray Cross-Correlation Function (CCF) shows the optical band lagging the X-ray by $~$0.5
s, but with a dip 2-5 seconds in advance of the X-rays (K01); 
(b) the correlation between X-ray and optical light curves 
appears to have timescale-invariant properties: 
the X-ray/optical CCF maintains a similar, but rescaled, shape on
timescales ranging at least from 0.1 s to few tens of sec (M03);
(c) the correlation does not appear to be  triggered by
 a single type of event (dip or flare) in the light curves; instead, as was
 shown by M03, optical and X-ray  fluctuations of very different shapes, amplitudes
 and timescales are correlated in a similar way, such that 
the optical light curve is related to the time derivative of the X-ray
one.
Indeed, in the range of timescales where the coherence is maximum,
the optical/X-ray phase lag are close to $\pi/2$, indicating that the
two lightcurves are related through a differential relation.
Namely, if the optical variability is representative 
of fluctuations in the jet power output  $P_{\rm j}$, 
the data suggest that the jet power scales roughly like $P_{\rm j} \propto
-\frac{dP_{\rm x}}{dt}$, where $P_{\rm x}$ is the X-ray power.

Malzac, Merloni \& Fabian (2004, hereafter MMF04) have shown that the
complex X-ray/optical correlations could be understood in terms of an
energy reservoir model.  In this picture, it is assumed that large
amounts of accretion power are stored in the accretion flow before
being channeled either into the jet  or into particle acceleration/ heating in the
comptonising region responsible for the X-rays. The optical variability is produced mainly from 
synchrotron emission in the inner part of the jet at distances of a few thousands
 gravitational radii from the hole. It is assumed  that  at any time the optical flux $O_{pt}$ (resp. X-ray flux)
  scales like the jet power $P_{\rm j}$ ( plasma heating power $P_{\rm x}$).
 MMF04 have developed
a time dependent model which is complicated in operation and
behaviour. However, its essence can be understood using a simple
analogue: Consider a tall water tank with an input pipe and two output
pipes, one of which is much smaller than the other. The larger output
pipe has a tap on it. The flow in the input pipe represents the power
injected in the reservoir $P_{\rm i}$, that in the small output pipe
the X-ray power $P_{\rm x}$ and in the large output pipe the jet power
$P_{\rm j}$.  If the system is left alone the water level rises until
the pressure causes $P_{\rm i}=P_{\rm j}+P_{\rm x}$.  Now consider
what happens when the tap is opened more, causing $P_{\rm j}$ to
rise. The water level and pressure (proportional to $E$) drop causing
$P_{\rm x}$ to reduce. If the tap is then partly closed, the water
level rises, $P_{\rm j}$ decreases and $P_{\rm x}$ increases. The rate
$P_{\rm x}$ depends upon the past history, or integral of $P_{\rm
j}$. Identifying the optical flux as a marker of $P_{\rm j}$ and the
X-ray flux as a marker of $P_{\rm x}$ we obtain the basic behaviour
seen in XTE\,J1118+480.  In the real situation, we envisage that the
variations in the tap are stochastically controlled by a shot noise
process. There are also stochastically-controlled taps on the input
and other output pipes as well. The overall behaviour is therefore
complex. This simple model is largely independent of the physical nature of the energy
reservoir. In a real accretion flow, the reservoir could take the form
of either electromagnetic energy stored in the X-ray emitting region,
or thermal (hot protons) or turbulent motions. The material in the
disc could also constitute a reservoir of gravitational or rotational
energy behaving as described above.
MMF04 show that the observed complex rapid optical X-ray correlation  described above and in  particular the differential relation between jet power (i.e. optical flux) and X-ray flux, can be explained by this  relatively simple basic model 
involving several energy flows and an energy reservoir. 
A strong requirement of the model is that the jet power should be at least a few times larger than the X-ray power.  This fact seems to be independently supported by  recent  results of on the study of the X-ray radio correlation in black holes and neutron stars indicating that for  black holes in the LHS  the jet power dominates over the X-ray emission (K\"ording et al. 2006).
\section{Conclusions}
\label{sec:conc}

The rich phenomenology of the connection between accretion and ejection is now quite well established.   Depending on luminosity different X-ray spectral states are observed suggesting that the nature and geometry of the corona depends on mass accretion rate. 
The observations of a correlation between the X-ray and radio band, indicates that the Compton corona is intimately related to the formation of compact radio jets and probably constitutes the base of the jet.
Those correlations occur during the evolution of the X-ray luminosity of the sources during outburst  lasting typically from weeks to months. But the coupling between  disc and jet can occur on times scales of hours and less during state transitions. I discussed an INTEGRAL observation of Cygnus X-1 during a mini-state transition. This observation suggests that during the state transition the jet power is unrelated to the fluctuations of the luminosity of the corona but strongly anti-correlates with the standard thin disc luminosity.  Finally the source XTE J1118+480 in the LHS, shows complicated correlations between the optical and X-ray band which are most likely a signature of coupling between the corona  and jet on time-scales of seconds or even less. 

Of course there are still many issues that  remain unsolved.  The geometry of the accretion flow in the hard state is still debated (accretion disc coronae versus hot accretion disc). The connection between the hot comptonising plasma  the compact jet is not robustly understood, neither are the mechanisms triggering state transitions. Finally as it is suspected for XTE J1118+480, many of the characteristics of the  fast X-ray variability of accreting black holes could be associated to some form of disc jet coupling and remain essentially mysterious.





\end{document}